\documentclass[sigconf, nonacm]{acmart}
\AtBeginDocument{%
  }
\usepackage{titlesec}

\titlespacing*{\section}
  {0pt}{1.5ex}{0.8ex}

\titlespacing*{\subsection}
  {0pt}{0.8ex}{0.5ex}
  
\begin{document}
\sloppy

\title{A Structured Review of Fixed and Multimodal Sensing Techniques for Bat Monitoring}

\author{Erwei He }
\email{erweihe@umich.edu}
\affiliation{%
  \institution{University of Michigan}
  \city{Ann Arbor}
  \state{Michigan}
  \country{USA}
}

\author{Sravya Ganti }
\email{sravya @umich.edu}
\affiliation{%
  \institution{University of Michigan}
  \city{Ann Arbor}
  \state{Michigan}
  \country{USA}
}

\author{Maatla Sefawe }
\email{maatla@umich.edu}
\affiliation{%
  \institution{University of Michigan}
  \city{Ann Arbor}
  \state{Michigan}
  \country{USA}
}

\author{Julianna Segalla }
\email{jsegalla @umich.edu}
\affiliation{%
  \institution{University of Michigan}
  \city{Ann Arbor}
  \state{Michigan}
  \country{USA}
}

\author{Isaac Tourner }
\email{itourner@umich.edu}
\affiliation{%
  \institution{University of Michigan}
  \city{Ann Arbor}
  \state{Michigan}
  \country{USA}
}

\author{Julia Gersey}
\email{gersey@umich.edu}
\affiliation{%
  \institution{University of Michigan}
  \city{Ann Arbor}
  \state{Michigan}
  \country{USA}
}

\renewcommand{\shortauthors}{y et al.}

\maketitle


\section*{Abstract}
Effective monitoring of mobile animal populations is crucial for ecological research, wildlife management, and agricultural applications. Monitoring of bats specifically can help understand the spread of disease as well as shine light on bat migration patterns, population dynamics, and the impacts of environmental changes on bat colonies. Fixed sensing modalities, such as infrared sensors, cameras, radar, and acoustic detectors, play a pivotal role in tracking and understanding animal behavior. This survey goes over context-informing details about bat biology, and then  reviews these fixed sensing modalities, discussing the unique challenges and contributions of each approach. We highlight the coverage, applications, accuracy, and limitations associated with each of these sensing modalities. By synthesizing recent advances, we provide a comprehensive overview to guide future research in this area. 
\section{Introduction}

Bats are among the most ecologically and economically significant mammalian groups, providing insect suppression, pollination, seed dispersal, and nutrient cycling across diverse biomes. These services create measurable downstream benefits for agricultural productivity and ecosystem stability, reinforcing the importance of bat conservation and management strategies \cite{RamirezFrancel2022BatsServices}\cite{TuneuCorral2023PestSuppression}.

At the same time, many bat populations face compounding pressures from habitat change, climate shifts, and emerging infectious disease. White-nose syndrome has caused severe mortality and regional population collapses in North American hibernating species, illustrating how rapidly disease can destabilize bat communities and therefore amplify conservation urgency \cite{Frick2010EmergingDisease}. 

Effective monitoring becomes critical for understanding population dynamics, migration, and behavioral responses to environmental change. Fixed sensing modalities including acoustic detectors, cameras, infrared systems, and radar offer noninvasive approaches for tracking bats and related species across spatial and temporal scales, while enabling long-term and repeatable deployments that reduce the need for continuous in-person sampling.

However, monitoring bats remains technically and ecologically challenging. Their small body size, fast and agile flight, nocturnal activity, and reliance on ultrasonic echolocation produce detection biases that vary by criteria such as species, habitat, and weather conditions. These constraints motivate sensing strategies that can either capture echolocation calls reliably or complement acoustics with optical, infrared, or radar-based observations. Because different modalities trade off coverage, species-level resolution, infrastructure requirements, and environmental robustness, a comparative synthesis is needed to guide modality selection and multimodal system design for bat monitoring.

In this survey, we first summarize key biological and behavioral features of bats that shape sensing requirements. We then review the landscape of fixed sensing technologies used in bat research and management including acoustic detection, camera-based computer vision, infrared sensing, and radar, highlighting their coverage characteristics, accuracy drivers, representative deployments, and limitations. Finally, we identify cross-cutting design challenges and outline opportunities for multimodal fusion and edge-enabled analytics to support scalable, low-disturbance bat monitoring.

\section{Biology of Bats}

Bats represent one of the most diverse and broadly distributed mammalian lineages, occupying an exceptionally wide range of ecological niches. Their roles as insect predators, pollinators, and seed dispersers position bats as sensitive bioindicators of ecosystem change, with population trends reflecting shifts in habitat quality, climate, and land-use patterns \cite{RamirezFrancel2022BatsServices}. 

\subsection{Nocturnality, Flight, and Roosting Ecology}
Most bat species are nocturnal and rely on roosts such as caves, tree hollows, rock crevices, bridges, and human-built structures. Roost choice and emergence timing are strongly linked to thermoregulation, predation risk, and reproductive state, causing predictable temporal patterns in flight activity that fixed sensors can exploit. The spatial bottlenecks created by roost entrances also motivate sensing designs focused on emergence counting and corridor-based monitoring.

\subsection{Echolocation and Sensory Specialization}
A defining feature of most insect-eating bats is echolocation, the production of ultrasonic calls and interpretation of returning echoes to navigate and locate prey. Call frequencies typically exceed 20 kHz and can reach much higher bands depending on species and behavioral context, enabling precise object detection but also imposing strict requirements on acoustic hardware bandwidth and sensitivity \cite{Jones2005Echolocation}. 

Echolocation call structure varies substantially across species. Species adapted to open spaces often emit lower-frequency, higher-intensity calls that propagate farther, whereas clutter-adapted species may use higher-frequency, lower-intensity signals optimized for fine spatial resolution in complex environments. This diversity directly affects detectability in passive acoustic surveys and introduces species-specific bias in fixed acoustic monitoring. These biological constraints also help explain why multimodal approaches, such as pairing acoustics with infrared or radar, can improve confidence in presence estimates and emergence counts.

\subsection{Movement Ecology and Migration}
Many bat species undertake seasonal movements ranging from local commuting to long-distance migration. These flights can be influenced by weather, landscape structure, and resource availability, generating spatiotemporal patterns that may be difficult to capture using local acoustic sensors alone. Radar studies, including analyses of weather-radar signatures, have demonstrated the feasibility of tracking large-scale nightly emergence and dispersal patterns in species such as Brazilian free-tailed bats, providing colony-scale and regional-scale movement insights that complement ground-based modalities \cite{HornKunz2008NEXRAD}.

\subsection{Disease Relevance to Monitoring}
Disease is a major driver of contemporary bat conservation. White-nose syndrome, caused by the fungal pathogen Pseudogymnoascus destructans, disrupts hibernation physiology and has led to precipitous population declines in several North American species \cite{Frick2010EmergingDisease}. Monitoring strategies that can detect changes in roost occupancy, emergence magnitude, and seasonal activity, therefore, play an important role in documenting disease impacts and evaluating intervention success. 

\subsection{Implications for Fixed Sensing Design}
Collectively, bat biology implies several concrete constraints for fixed sensing systems. First, nocturnality and low-light flight environments favor acoustic and infrared approaches over visible-light-only imaging. Second, species-specific echolocation structure and atmospheric effects on ultrasound propagation impose stringent requirements on microphone selection, placement, and weather-aware sampling protocols. Third, roosting behavior creates high-value monitoring points (i.e. entrances, flyways) that can support accurate emergence estimation with appropriately placed acoustic arrays, thermal sensors, or beam-break systems. Finally, the existence of migration and colony-scale mass emergences motivates radar as a complementary modality for coverage beyond local acoustic range.

These biological realities underscore why no single sensing approach is universally sufficient, and they motivate the modality-by-modality evaluation presented in the remainder of this survey.

\section{Acoustic Detection}
Beginning in the early 1970s, acoustic monitoring has become a widely used and effective noninvasive method for studying bat populations\cite{walters_challenges_2013}\cite{roemer_current_2025}. This approach is well-suited for bat species as they rely on echolocation, or the emission of ultrasonic calls and interpretation of the returning echoes, to navigate their surroundings and locate prey in the dark\cite{jones_echolocation_2005}\cite{walters_challenges_2013}. Bats themselves serve as valuable indicators of biodiversity, as their global distribution and behaviors can reveal important information about ecosystem health, emerging disease risks, habitat change, and the impacts of climate change\cite{macswiney_g_what_2008}\cite{walters_challenges_2013}. Given their ecological importance, effective monitoring is essential; however, acoustic methods come with their own challenges. This section reviews coverage, accuracy, ecological impacts, and limitations associated with acoustic detection in fixed bat-sensing systems.

\subsection{Active and Passive Acoustic Detection}
Acoustic detection generally takes two forms: active and passive. Active acoustics generate a sound wave and analyze the returning echoes\cite{mellinger_overview_nodate}\cite{mann_active_2008}. In contrast, passive detection does not produce an outgoing signal; instead, it records sounds naturally produced by the environment and the animals within it \cite{mellinger_overview_nodate}\cite{marques_estimating_2013}\cite{mann_active_2008}.

Both approaches have been widely used to monitor a variety of animal groups. Passive acoustic monitoring has been extensively applied to marine mammals such as whales and dolphins\cite{mellinger_overview_nodate}, as well as to fish\cite{mann_active_2008}, birds\cite{ntalampiras_acoustic_2021}, and forest elephants\cite{wrege_acoustic_2017}. Active acoustic detection is also common in marine ecosystems, particularly for studying fish, squids, and shrimp\cite{mellinger_overview_nodate} \cite{mann_active_2008}.

Passive acoustic detection offers several advantages over traditional visual surveys and physical trapping. It performs well in dark or light-limited conditions and is less affected by weather\cite{marques_estimating_2013} \cite{heinicke_assessing_2015}. It also enables simultaneous monitoring of multiple taxonomic groups and reduces human interaction in the study area to only installation and maintenance, thereby minimizing disturbance to wildlife and their habitat\cite{heinicke_assessing_2015}.

In previous work, passive acoustic detection has been used extensively to study bat populations\cite{aodha_bat_2018}\cite{ahlen_use_1999}\cite{milchram_estimating_2020}\cite{macswiney_g_what_2008}. This is largely because most bat species rely on echolocation, meaning much can be inferred about their behavior and activity simply by analyzing their calls. Numerous studies have explored how acoustically cluttered environments affect bat behavior\cite{allen_effect_2021}\cite{arlettaz_effect_2001}\cite{amichai_calling_2015}. These works have shown that bats become less effective at locating prey in cluttered conditions\cite{arlettaz_effect_2001}. To cope with such clutter, or acoustic jamming, bats adjust their echolocation calls to improve their signal-to-noise ratio\cite{amichai_calling_2015}. However, these adjustments do not fully overcome the effects of clutter. Sonar-based object discrimination often becomes impaired, although in some cases clutter can enhance discrimination abilities\cite{allen_effect_2021}.

Given these considerations, passive acoustic detection is often preferred when the goal is to study bat behavior while minimizing disturbance to their natural activity and habitat. It is also well-suited to bats’ nocturnal behavior and the low-light environments in which they forage and navigate. For this reason, the remainder of this section will focus mainly on passive acoustic detection.

\subsection{Coverage and Effective Acoustic Detection}
Acoustic detection is widely used to monitor bats, supporting objectives such as estimating activity trends, documenting foraging habits, and identifying species. Its effectiveness depends on both ecological and technical factors that stem from species-specific characteristics and detector selection.

\subsubsection{Biological Factors}
Different bat families and species produce echolocation calls with distinct characteristics, particularly in frequency and intensity, and these traits directly influence how detectable they are through acoustic monitoring. Bats that emit higher-frequency, low-intensity calls are generally less likely to be detected than species producing lower-frequency, high-intensity calls \cite{adams_you_2012}\cite{macswiney_g_what_2008}. Higher-frequency signals attenuate more rapidly, reducing detection range and potentially leading to under representation of certain species in acoustic surveys\cite{adams_you_2012}\cite{murray_characterization_2009}.A well-documented example is the Phyllostomid family, which remains difficult to classify due to both low call intensity and limited inter-specific variation\cite{macswiney_g_what_2008}. Flying behavior also plays a role in detectability. Many bats do not typically fly below two meters, meaning that placing microphones at this height may reduce detection success\cite{frick_acoustic_2013}. However, some species such as trawling bats or aerial insectivores like Pteronotus parnelli regularly fly at lower heights\cite{frick_acoustic_2013}.

Given these considerations, it is essential to account for the biological characteristics of the species being surveyed. When these characteristics are unknown, survey design should allow for a wide range of call intensities and flight heights to ensure the most accurate and representative results.

\subsubsection{Technical and Hardware Factors}
The technical characteristics of a detector strongly influence acoustic sampling quality. Numerous commercial systems exist, such as the Avisoft UltraSoundGate 116 CM16/CMPA, AnaBat SD2, Batcorder 2.0, Batlogger, and Son Meter SM2BAT. Choosing among these devices, or designing a custom detector, directly impacts the quality, precision, and amount of data collected\cite{adams_you_2012}. 

A major source of detection bias arises from the interaction between bat call properties and detector frequency response. As stated above, species producing low-intensity, high-frequency calls are inherently less detectable. This ecological bias is amplified when microphones have reduced sensitivity at upper frequencies, further lowering detection probability for these species\cite{adams_you_2012}. As a result, microphone performance must be evaluated relative to the call frequency range of the local bat community, which can span roughly 8–200 kHz\cite{adams_you_2012}\cite{brigham_bat_2004}.

Microphone sensitivity, directionality, and placement also shape detection range and airspace coverage. More sensitive microphones detect bats at greater distances, while detectors with omnidirectional elements have wider angular coverage but typically shorter range due to lower sensitivity\cite{adams_you_2012}. Orientation and weatherproofing methods used matters as well. Horizontally oriented microphones and certain weatherproofing designs have been shown to markedly reduce call counts and species detections\cite{britzke_effects_2010}. These effects illustrate that there is no universally optimal configuration; instead, equipment choices must align with expected call types, site conditions, and sampling goals.
Together, these hardware factors determine how effectively a detector samples the acoustic environment and must therefore be carefully matched to the species, frequencies, and habitats being surveyed.

\subsubsection{Methodological and Complementary Approaches}
When calls are distinct enough to separate species, acoustic surveys often detect more species than capture-based methods\cite{ofarrell_comparison_1999}. Still, many studies recommend pairing acoustic data with complementary approaches such as capture or visual confirmation to improve coverage and species-level accuracy\cite{ofarrell_comparison_1999}\cite{clement_accounting_2014}\cite{ahlen_use_1999}\cite{macswiney_g_what_2008}. So while acoustic detection can be powerful on its own, pairing it with a complementary approach can help strengthen results. These complementary methods may include cameras, computer vision techniques, infrared sensing, and radar. All of these approaches are discussed in detail in later sections of this paper. 

\subsection{Acoustic Detection Accuracy}
Accuracy in acoustic bat detection is influenced by several biological, technical, and environmental factors. Acoustic surveys often capture more activity than traditional methods, but they are most reliable when paired with capture techniques or visual confirmation, or another complementary approach\cite{ofarrell_comparison_1999}\cite{clement_accounting_2014}.

\subsubsection{Identification Errors and Computational Challenges in Acoustic Analysis}
A common analytic method used for acoustic surveys is occupancy models, which assume that misidentification does not occur. This assumption can lead to incorrect inferences about collected data\cite{clement_accounting_2014}. The similarity among bat species further increases the likelihood of misidentifications and false positives during acoustic surveys\cite{clement_accounting_2014}. As a result, there is a continued effort to develop more sophisticated analytic tools to improve identification accuracy. Recent approaches include discriminant function analysis, neural networks, and support vector machines\cite{clement_accounting_2014}\cite{preatoni-2005}\cite{britzke-2011}. It is important to consider the errors introduced by any analytic method chosen for an acoustic survey and work to minimize them so they do not significantly affect study conclusions\cite{clement_accounting_2014}.

In addition to analytic limitations, achieving high accuracy is challenged by the need to process increasingly large acoustic datasets. There is a growing need for open-source tools capable of detecting and classifying bat calls at scale\cite{aodha_bat_2018}. Many existing systems are commercial and focus primarily on species classification, but more emphasis is needed on accurately locating echolocation calls within noisy recordings\cite{aodha_bat_2018}. Advances in machine learning have contributed to improved detection performance, particularly through deep learning models using convolutional neural networks, which have shown significant gains in identifying search-phase echolocation calls in full-spectrum recordings\cite{aodha_bat_2018}. These approaches offer a promising direction for reducing false positives and increasing overall accuracy.

\subsubsection{Environmental Effects on Accuracy}
Echolocation calls are effected by both atmospheric and geometric attenuation\cite{goerlitz_weather_2018}. The speed of sound and atmospheric attenuation are influenced by temperature and humidity weather conditions, so variations in weather conditions can be a source of noise when measuring bat acoustic activity\cite{goerlitz_weather_2018}\cite{roemer_current_2025}. In addition, bats are known to avoid low temperatures and strong winds, so acoustic sampling should be performed during ideal weather conditions\cite{roemer_current_2025}. It is important to not make assumptions about the absence of a bat population if acoustic detection sampling has been done during higher temperatures and humidity\cite{borkin_influence_2023}. To be confident in results, surveys should be done for multiple nights in suitable conditions\cite{borkin_influence_2023}.

Overall, achieving reliable acoustic detection requires mitigating analytic and environmental sources of error through thoughtful survey design and robust processing tools. Complementary methods continue to play a key role in verifying detections and ensuring accurate species identification.

\subsection{Conclusions on Acoustic Detection for Bat Monitoring}
Overall, acoustic detection has become a critical tool for monitoring bat populations, offering a noninvasive way to assess activity, behavior, and ecological patterns across diverse environments. Its effectiveness, however, depends on accounting for biological variability in call structure, technical and hardware limitations, and environmental conditions that influence signal propagation. Advances in signal processing and machine learning have improved identification accuracy, yet misclassification and false positives remain important considerations. Because of these challenges, acoustic data are most reliable when integrated with complementary methods such as capture, visual confirmation, or additional sensor modalities. When applied thoughtfully, acoustic detection provides a strong foundation for generating accurate assessments of bat presence, activity, and community patterns.



\section{Cameras and Computer Vision} 
The integration of Camera Traps (CTs) with Computer Vision (CV) and Artificial Intelligence (AI) represents a fundamental shift in wildlife monitoring, moving away from fragmented, labor-intensive approaches toward scalable, automated environmental surveillance\cite{chalmers_removing_2023}.

Camera traps are remote devices that automatically capture still images or videos when triggered by animal movement, typically using Passive Infrared (PIR) sensors\cite{nazir_wiseeye_2017}. This technology provides a cost-effective, non-invasive method for sampling communities of mid-to large-terrestrial species, allowing for repeat surveys and continuous, long-term deployment in remote areas\cite{nazir_wiseeye_2017}\cite{kline_smartwilds_2025}\cite{lahoz-monfort_comprehensive_2021}.

\subsection{Applications and Data Collection}
CTs serve various conservation objectives by capturing visual documentation specifically structured for object detection, tracking, and fine-grained classification of wildlife species\cite{kline_smartwilds_2025}.

\subsubsection{Data Output}
Modern camera traps generate auditable, long-term data archives that are time-stamped and often geotagged\cite{nazir_wiseeye_2017}. Each image or video comes with metadata including timestamp and sometimes GPS coordinates or site IDs. This rich metadata can be leveraged for multi-sensor fusion studies. For instance, the SmartWilds dataset provides a synchronized collection of camera trap photos/videos, drone imagery, and bioacoustic recordings collected together in the field\cite{kline_smartwilds_2025}.

Such integrated datasets enable researchers to cross-validate detections across modalities (image, audio, etc.) and gain a more comprehensive view of an ecosystem. Camera trap datasets, when properly curated, become valuable resources for conservation planning and can be shared openly to accelerate research. Some platforms even allow users to upload camera trap images and apply AI models in the cloud, significantly reducing the time required to convert raw images into usable biodiversity information \cite{ahumada_wildlife_2020}.

\subsection{Hardware and Operational Challenges}
A crucial component of most camera traps is the PIR motion sensor, which detects animals by sensing changes in infrared radiation. While PIR sensors are energy-efficient and work both day and night, they introduce several limitations:
\subsubsection{Limited Sampling Area}
A camera trap only “sees” a narrow detection zone in front of its lens. It will detect animals only within a certain range and field of view. Animals outside the frame, stationary animals, or those moving just outside the range often go undetected.
\subsubsection{Environmental Sensitivity}
PIR sensitivity is highly dependent on temperature contrast, becoming ineffective above $35^\circ\text{C}$\cite{hobbs_improved_2017}. High sensitivity settings, necessary for capturing small or fast-moving animals, amplify the likelihood of generating false positives triggered by environmental disturbances like wind-blown vegetation, rain, or heat reflections \cite{hobbs_improved_2017}\cite{nazir_wiseeye_2017}.

\subsubsection{Trigger Speed}
The time between detection and image capture is critical. If the trigger is too slow, a camera may capture only the tail end of an animal or miss it entirely as it moves past.

\subsection{Thermal Infrared Imaging}
Thermal cameras are an example of passive thermography. These cameras detect animals and quantify their thermal state by emitting long-wave IR and subsequently producing a 2D temperature map. \cite{travain_infrared_2021}\cite{kadlecova_use_2024}\cite{wagner_thermal_2025}\cite{wongsaengchan_non-invasive_2023}\cite{madani_size_2025}

This type of camera enables the detection of nocturnal animals such as bats and is effective with long-range imaging and covert tracking. Thermal imaging has been used to support the analysis of movements paths, habitat use, and social interactions of wildlife. 
For related reasons, thermal cameras are also often used in nocturnal survey applications.

In a more granular use case, thermal imaging has been used to detect the health status of wildlife by quantifying localized temperature changes that might be markers of stress. An example is detection of a change in temperature around the eyes of an animal to conduct a non-invasive welfare assessment. \cite{hristov_applications_2008}\cite{wagner_thermal_2025}\cite{dos_santos_thermal_2025}\cite{liehrmann_enhancing_2024}

\subsection{Advancements in Computer Vision and AI}
The challenge of manually sorting, cataloging, and extracting data from the vast datasets generated by high-throughput camera traps necessitates the use of computational sophistication, particularly AI \cite{lahoz-monfort_comprehensive_2021}.

\subsubsection{Automated Data Processing}
Deep Learning (DL) has fundamentally transformed the extraction of ecological knowledge from large datasets. AI algorithms can swiftly identify and catalog wildlife from camera trap images, which accelerates data processing and enhances conservation planning. DL applications in this domain support tasks such as object detection, tracking, and species classification\cite{chalmers_removing_2023}\cite{wang_wb-yolo_2025}.

\subsubsection{Architectures and Tasks}
The input data for image DL systems resembles digital images, making Convolutional Neural Networks (CNNs) highly effective and dominant in classification tasks\cite{stowell_computational_2022}. Frameworks like YOLO-Behaviour are utilized for identifying visually distinct behaviors (such as frame-wise detections of behaviors in pigeons and zebras) from video recordings\cite{wang_wb-yolo_2025}\cite{kline_smartwilds_2025}\cite{chan_yolo-behaviour_2025}.
\subsubsection{Real-Time Processing and Edge AI}
DL models can be deployed using Edge AI (e.g., on dedicated hardware like Google's Coral microcontroller) directly on simple devices like camera traps, allowing them to run algorithms locally in real time\cite{gardiner_towards_2025}. This capability enables immediate data filtration and supports real-time applications\cite{iwane_real-time_nodate}.

The ability of CV techniques to rapidly process images and videos, combined with hardware innovations like edge computing and multi-modal integration, is crucial for turning the overwhelming volume of raw camera trap data into actionable ecological insights.
\section{Infrared Sensing}
The following section will discuss how infrared sensing can be used to collect information about mobile animals such as bats. The discussion will highlight how IR-based methods can detect, count and assess the behavior of these animals in a contact-free way. \cite{karp_detecting_2020}\cite{hristov_applications_2008}\cite{lathlean_edge_2017}\cite{mccafferty_value_2007}\cite{mccafferty_applications_2013}

The use of this sensing methodology is motivated by the need for non-invasive monitoring. It is also motivated by the simple fact that bats are nocturnal animals and low-light monitoring is essential for understanding their movement, activity budget, and state, a use case where visible-light cameras are impractical. \cite{rietz_drone-based_2023}\cite{wagner_thermal_2025}\cite{mccafferty_value_2007}\cite{mccafferty_applications_2013}

Passive infrared sensing (PIR) senses changes in ambient thermal radiation, and in contrast Active IR sensors emit infrared beams or structured light and detect interruptions. Common passive IR methods include thermal cameras and thermography. \cite{mota-rojas_thermal_2022}
Well-known Active IR methods include beam-break counters, IR LED-detector pairs, and IR-illuminated video. \cite{travain_infrared_2021}\cite{rafique_parametric_2012}\cite{yun_human_2014}

\subsection{Infrared Sensing for Bats}
To demonstrate the contextual link between IR sensing and bat monitoring, this section will describe the current landscape of IR sensing for bat monitoring and behavior analysis. IR-illuminated video is used to record bat emergence and behavior at cave or roost entrances. This type of video uses Active IR principles by emitting near-IR light to trigger a camcorder to record activity. \cite{whiting_can_2022}\cite{jaffe_evaluating_2025}\cite{roby_testing_2025}

IR cameras and recorders are currently regarded as the most accurate method for exit counts at bat roosts, especially when the animals emerge after dark. \cite{krivek_counting_2023}\cite{lefevre_automated_2025}\cite{noauthor_docdm590789_2012}

In addition to solely using IR as the monitoring strategy, it is also common to use IR alongside other sensing methodologies covered in this paper, including acoustic monitoring, to collect information about bat swarming and roost occupancy dynamics. \cite{kloepper_estimating_2016}\cite{whiting_can_2022}\cite{jaffe_evaluating_2025}\cite{laurenzi_vocal_2025}

\subsection{Passive Infrared Sensing}
Passive IR sensors are low power devices that receive and measure external IR emissions as input without emitting energy themselves. These sensors collect information about wildlife and human behavior by detecting changes in infrared radiation (heat) that are naturally emitted by objects within their detection area. Passive IR sensors detect temperature changes with a pyroelectric material sensitive to IR. They also often use a Fresnel lens, which focuses incoming radiation onto the pyroelectric sensors. All objects above absolute zero temperature emit some level of electromagnetic radiation. Therefore, when an animal passes in front of the sensor, a change is detected, indicating a motion event \cite{caniou_passive_1999}\cite{welbourne_how_2016}.

\subsubsection{Passive Infrared Motion Sensors}
PIR motion sensors are used in small-animal activity monitoring because these sensors can be quite low-power and function without disturbing animals. Commercial PIR sensors for this application use AC-coupled detection to read and quantify position changes or assess the gross activity of small animals in a cage. \cite{rafique_parametric_2012}\cite{thang_setup_2021}\cite{yun_human_2014}\cite{singh_low-cost_2019}\cite{gabloffsky_establishment_2025}\cite{travain_infrared_2021}

These motion sensors are also used in detection of large animals such as humans. PIR-based motion detection and occupancy sensors can be used to sense household occupancy and human behavior patterns. Some are even immune to false positives from pets, an example of PIR sensors being able to filter out unnecessary information for the current application. \cite{yun_human_2014}\cite{aldalahmeh_enhanced-range_2016}

\subsubsection{Differential and Multi-Element Designs}
Differential IR sensing employs PIR elements while adding additional functionality to cancel out background radiation and provide a more robust detection of moving objects. Differential IR sensors combine multiple PIR sensing zones that are wired in opposition, so that when an animal moves across the field of detection, the detection level at one zone is subtracted from the other. This difference when generated creates a pulse that signifies a motion event. This can be more robust than a single zone PIR sensor that can be triggered by slower temperature changes. \cite{shine_occupancy_2022}\cite{noauthor_infrared_2025}

This is taken a step further with sectorized lenses. The Fresnel lens that is used to focus IR onto the pyroelectric material in a PIR sensor can be segmented to focus IR waves from a larger field into the sensor, making it more sensitive to motion. \cite{noauthor_what_2025}

Another general principle of effective motion detection using PIR is the combination of this sensing modality with others and the requirement of a detection across all modes to verify a motion event. PIR is often combined with another independent modality such as microwave Doppler radar or acoustic detection. \cite{noauthor_dual_2025} 

This is also helpful when creating pet-immune motion detection systems. The use of microwave signatures or sound patterns can help filter out thermal detection that isn't useful for the use case. This can be extended to environments with multiple kinds of wildlife but only one species of interest. \cite{noauthor_dual_2025}\cite{noauthor_honeywell_2025}

\subsubsection{Limitations of Passive Infrared Sensing}
A persistent limitation of PIR sensing is its vulnerability to false positives caused by environmental temperature changes, or hot surfaces. \cite{travain_infrared_2021}\cite{kadlecova_use_2024}\cite{}\cite{mccafferty_value_2007}

Additionally, PIR sensors in particular are limited by occlusion and line-of-sight issues. Cluttered and complex habitats can reduce count accuracy. This problem is exacerbated by small and tightly clustered animals, that may produce overlapping thermal signatures. This is often addressed by employing additional modalities to help conduct individual-level behavioral assessment. \cite{wagner_thermal_2025}\cite{hristov_applications_2008}\cite{lathlean_edge_2017}\cite{mota-rojas_thermal_2022}\cite{madani_size_2025}

\subsection{Active Infrared Sensing}
Active infrared sensors actively emit infrared light in contrast to passively detecting infrared like PIR sensor systems. They commonly use IR LEDs or laser diodes to do so and then collect information based on how the light is received, or interrupted by objects in the environment. \cite{noauthor_infrared_2024}

The two most common architectures are break-beam and reflective. In break-beam systems, the emitter and receiver face each other and are placed some distance from each other. When an object moves in between them, it is detected as a beam break. In reflective systems, emitters and receivers are on the same side and when the beam reflects off of an object the sensor measures the reflected IR. A common embedded systems application is a time of flight (TOF) sensor. \cite{krivek_counting_2023}\cite{lefevre_automated_2025}\cite{koch_how_2023}

\subsubsection{Infrared Beam-Break and Gate Systems}
Active IR beam counters when used for bat sensing are often strategically placed at roost exits and corridors. The animal passage interrupts the emitted IR beams and thus generates timing or count information. This information can then be used to inform bat emergence estimates. \cite{noauthor_infrared_2025}\cite{noauthor_batcounter_2025}

Dedicated bat counter devices often use rectangular IR beam gates to count bats as they pass through defined apertures. These devices are a quick, automated, field-deployable solution for bat counting.\cite{krivek_counting_2023}\cite{welbourne_how_2016}

\subsubsection{Active Infrared Sensing in Lab and Husbandry Settings}
In lab settings, active IR sensing is often used in mazes to detect important events in rodent and other small animal behavior. These events are detected by beam interruption in line with previously discussed active IR sensing methods.\cite{noauthor_infrared_2025-3}\cite{madani_size_2025}\cite{noauthor_ad-2_2025}

Similar technology is used in cage-top motion detectors. This helps provide information about animal circadian rhythm, activity budget, and drug effects. \cite{gabloffsky_establishment_2025}\cite{singh_low-cost_2019}\cite{travain_infrared_2021}

In husbandry or farm applications, active IR and IR illuminated cameras can be used to detect animals in fields, monitor animal body temperature, and prevent wildlife injuries. These animals often move more slowly and are not airborne, but it is evident how the information gleaned from this sensing methodology can be useful for bat sensing. \cite{lathlean_edge_2017}\cite{kadlecova_use_2024}\cite{wagner_thermal_2025}

\subsubsection{Limitations of Active Infrared Sensing}
A limitation of active infrared sensing is the propensity for count errors when relying on beam-break encounters. When multiple animals pass across an IR beam simultaneously, which is evidently an issue with bat swarming, the count can lose accuracy. If animals reverse direction while crossing the beam that can also cause counting issues, which affords a difficulty for aerial animal detection. \cite{krivek_counting_2023}\cite{kloepper_estimating_2016}\cite{lefevre_automated_2025}\cite{noauthor_docdm-131260_2012}

Robustness to the environment has also been an issue for active infrared sensing. Outdoor environments are vulnerable to rain, fog, vegetation movement, and insects, all of which can cause breaks or attenuation to received signals. \cite{pawar_motion_2018}\cite{holroyd_best_2023}\cite{koch_how_2023}

\section{Radar-Based Sensing for Wildlife and Bat Monitoring}

\subsection{Fundamental Concepts of Radar}
Radio detection and ranging (radar) works by sending electromagnetic waves into a desired region and analyzing the reflections from objects to determine their presence, position, and motion. When a radar receives a signal, a portion of it is reflected back to the radar receiver, producing an echo from which range, direction, and velocity are inferred. The echo-based detection principle is foundational: a received signal encodes the expected range of a target through the round-trip signal delay, denoted by \(D = cT/2\), and encodes the target's direction through the antenna's azimuth and beam angle. The integral components of the system for the process: an RF energy source, an antenna system for transmitting and receiving waves, a duplexer or switching between transmit/receive, an amplified receiver to amplify weak echoes, and a signal-processing system to extract readable target information from changes in amplitude, phase, and frequency.\cite{anjaneyulu2017mini}

Current radar systems can cover most of the radio spectrum, from UHF at hundreds of MHz to millimeter-wave systems at 24–79 GHz. Low-frequency bands provide long-range detection with larger antennas, but at lower resolution, whereas higher-frequency bands achieve high spatial resolution for small targets—a key consideration for wildlife sensing—though limited in maximum range. X-band (8–12 GHz), Ku-band (12–18 GHz), and K-band (18–26.5 GHz) offer a good resolution and reasonable propagation losses and are widely used in surveillance, environmental monitoring, and compact sensing platforms. Radar technology excels at operating in the dark, under fog, and in high-noise conditions; therefore, it is most appropriate for nighttime or fast-moving animal tracking, where optical or sonic methods may not be effective.\cite{anritsu2023basics}
Two dominant types of radar systems are pulse radar and CW (continuous-wave) or frequency-modulated continuous-wave (FMCW) radar. Pulse radar uses very short, high-power pulses and determines range from the time lag between transmission and echo reception. Their performance is driven by pulse parameters: while pulse width is the resolution of the range and the shortest detection distance, pulse repetition frequency defines the maximum unambiguous range. Methods of pulse compression enable processing of long pulses (which improve the signal-to-noise ratio), and their effects will be apparent; thus, sensitive sound is obtained for both long-range monitoring and high-res sensing. In contrast, CW and FMCW radar transmit continuously and measure target velocity using Doppler frequency shifts. Frequency-modulated continuous-wave (FMCW) modules can encode range information via frequency modulation and provide high sensitivity with minimal peak transmission power. Because of these characteristics, FMCW radar becomes attractive for compact, low-power sensing systems with applications in environmental and biological monitoring.\cite{bakare2022comprehensive}\cite{anritsu2023basics}

Radar information extraction is based on how the amplitude, phase, and frequency of the returned waveform are changed by the target. Changes in amplitude with position or frequency reveal information about target size and physical characteristics, and phase differences over time or frequency indicate velocity and range. The Doppler effect due to relative motion between the radar and the target corresponds to a frequency modulation, allowing the calculation of velocity, which is critical for characterizing flight dynamics in biological sensing. As small targets like bats have low radar cross-section, successful detection requires selecting the frequency, antenna gain, waveform parameters, and high receiver sensitivity and processing. These relationships are well documented in theory by the classical radar equation, which describes the received power as a function of transmitted power, antenna gains, wavelength, radar cross section, and range, and underscores the importance of low-loss system design for monitoring small or distant organisms.\cite{bakare2022comprehensive}

In summary, the basic physics of radar, including time-of-flight range, Doppler-based velocity calculation, and angle measurement using antenna beam steering, offers an end-to-end noncontact sensing solution. In combination with its ability to operate in low-visibility and night-time modes, radar may be used as a companion to other unobtrusive wildlife sensing modalities, namely acoustic or infrared detection. Its mature nature, along with the scalability and variety of behavior across frequency bands, is desirable and suitable for use in long-range bat activity monitoring, flight trajectory maps, and monitoring behaviors in contexts where conventional sensing is constrained.

\subsection{Motivating Radar for Bat Monitoring}
Monitoring bats presents long-standing challenges due to their nocturnal behavior, small body size, fast, agile flight, and tendency to occupy large, three-dimensional spaces that are difficult to observe directly. While acoustic devices of all sorts are still widely adopted, the volumes at which they detect are usually quite limited and depend on a particular frequency; for example, in many species, calls fade almost quickly, and in complex environments (for example, wind turbines), acoustic detection instruments may only see a fraction of the actual operating volume. As wind power expands further into the field, with taller turbines and larger areas whipped up by the wind, ecological monitoring methods must be scaled accordingly. These limitations have sparked considerable interest in radar as a complementary, high-coverage sensing modality for wildlife and bat monitoring. 

Radar has long been deployed in aeroecology, used for decades to study bird and insect migration, prior to its newfound potential for bat research being recognized more widely. National Doppler networks and other weather radars gather large, continuous ambient data and have revealed spatiotemporal patterns of animal movement across continents. These networks provide information on migration timing, altitude distributions, mass-movement events, and reactions to weather systems that cannot reasonably be captured by a single tracking device. Operational weather radars, as underscored by continental-scale initiatives, offer strong opportunities to detect these biological signals, provided that appropriate algorithms and classification techniques are developed to separate animals from meteorological echoes \cite{shamoun2014continental}.

 At the regional and local scales, marine and tracking radars, and modified X-band systems dedicated to ecological radar have demonstrated the ability to detect individual bats in flight. Preliminary migration studies conducted with vertically oriented (vertical) marine radars recorded nocturnal bird and bat passage rates, altitudes, and where they overlapped with wind turbine rotor-swept zones. For example, studies at proposed wind energy parks demonstrated that radar was used to assess bird and bat migration speed and intensity, flight height relative to turbine buildings, and temporal motion during the high migration season \cite{mabee2005radar}. This highlighted the role of radar in assessing collision risk when acoustic or visual cues are lacking. 

Recent innovations have further confirmed radar’s relevance in bat-specific identification. Newer X-band pulse radars, coupled with high-resolution image processing and a network of satellite or auxiliary sensors, can track individual bats near structures such as windmills. For instance, a recent study showed that pulse radar and acoustic detectors together can directly verify radar-determined targets and associate radar tracks with bat species, enabling the authors to more accurately attribute radar tracks to specified bat species. Studies of this type have indicated that radar can quantify bat velocity, for example, and estimate in-flight radar size and detection volumes that may well exceed the range observed by acoustic sensors, thereby enabling the detection of bats with known morphological and behavioral profiles \cite{krapivnitckaia2024detection}. These findings substantiate radar’s ability to effectively record genuine bat behavior at scales equivalent to turbine-level risk estimation. 

Radar signal analysis methodologies, initially developed for insects—particularly micro-Doppler spectral analysis—have enabled high-frequency radars to extract signals from wingbeat patterns and motion dynamics. Although their specific applications had already been demonstrated in insects, this work clearly shows that radar can also measure minuscule flight features and suggests that the system is applicable to large insects and birds, including wingbeat frequency and modulation profiles, as a basis for bat separation \cite{diyap2022microdoppler}. 

Weather-radar–based approaches have also seen a dramatic expansion alongside local-scale radars. This includes using machine learning to characterize biological echoes generated by networked S-band radars, enabling near-real-time identification of bat emergences at landscape and continental scales. For example, neural network classifiers trained on NEXRAD data can accurately characterize free-tailed bats and generate continuous spatiotemporal maps of their presence. These systems enable scientists and managers to identify seasonal movements, emergence timing, and large-scale distribution characteristics beyond what can be inferred from ground-based acoustic or thermal sensors alone \cite{lee2024bats}. 

Together, this body of work shows that radar is becoming a necessity in wildlife and bat monitoring. Due to its ability to record movement patterns across large spatial scales, identify species in complex environments, overcome acoustic limitations, and interface with machine learning pipelines, radar is becoming an important tool in current aeroecology. Driven by extensive wind-energy production, increasing conservation requirements, and growing interest in cross-disciplinary multimodal sensing, monitoring bats using radar becomes an appropriate strategy for both scientifically rigorous and operationally scalable understanding and mitigation of bat population impacts.

\subsection{Radar Methodology}
Radar monitoring of bats and other volant wildlife is based on coordinated developments in radar apparatus, signal acquisition techniques, micro-Doppler analysis, weather-radar bio-extraction pipelines, classification protocols, and multi-sensor integration. Collectively, these ten main studies will contribute to understanding how biological targets can be detected, characterized, classified, and forecast spatially (from local sensing radars to continental weather radar networks).
\subsubsection{Radar Platforms and Micro-Doppler Extraction}
Complex, high-frequency coherent radars are used to quantify wingbeat-induced micro-Doppler signatures in detail. W-band (94–95 GHz) radars capture periodic amplitude and phase changes caused by diminutive insects, underscoring the importance of micro-Doppler-related micromovement in monitoring millimeter-wave systems \cite{wang2017micro}. Equally controlled experiments at K-band and W-band demonstrate that drones and avians can generate distinct micro-Doppler spectrograms due to their varying wingbeat frequencies, modulation depths, and harmonic structures \cite{rahman2018radar}. Altogether, these papers create methodological guidelines for acquiring wingbeat attributes to discriminate bats from birds or insects: coherent radar operation, high sampling rate, time-frequency analysis (STFT/cepstrograms), and micro-motion feature extraction.
\subsubsection{Weather Radar Signal Acquisition and Biological Echo Extraction}
A number of C-band and S-band operating weather radars have been widely utilized for large-scale bio-monitoring. Basic approaches for determining biological signals from Doppler weather radar volumes were established by Van Gasteren et al., who defined criteria for separating birds from meteorological echoes by thresholding reflectivity levels, integrating with elevation scans, and comparing with a dedicated wildlife radar \cite{vangasteren2008extracting}. Building on this approach, Nilsson et al. collected bird movements at regional scales by combining C-band radar measurements from multiple instruments, demonstrating that multi-radar mosaics are viable for estimating migration traffic patterns and movements \cite{nilsson2018quantifying}. Dokter et al. developed automated algorithms to find flight altitudes based on vertical density profiles and Velocity–Azimuth Display (VAD) fitting applied to networks of operational weather radars \cite{dokter2011flight}. These approaches—altitude profiling, reflectivity integration, and Doppler-based motion estimation—establish the fundamental processing pipeline for extracting biological movement parameters from atmospheric radar networks.
\subsubsection{Clutter Suppression, Feature Segmentation, and Classification}
Bioscatter and non-bioscatter are crucial methods for biological conservation applications that require reliable differentiation. Hüppop et al. detail the methodological hurdles of clutter suppression, including the suppression of ground returns, anomalous propagation, sea clutter, and precipitation, for example, using static clutter maps, Doppler filtering, and polarimetric signatures \cite{huppop2019radar}. At the algorithmic level, Williams et al. propose a radar-based detection system to reduce aviation hazards that uses multi-parameter feature extraction (reflectivity, velocity, spectrum width) and supervised classification to detect avian targets in real time \cite{williams2018aviation}. The works of these approaches informed later wildlife-monitoring frameworks, which emphasize rigorous quality control, pixel-level feature segmentation, and a biological "objects" of detection prior to any ecological considerations, and a high level of biological material knowledge. Smith et al. show how a combination of radar–visual monitoring for radar detection at an offshore wind farm, which confirms and augments classification confidence by complementing reflectivity measurements with manual behavior measurements, enables the establishment of radar detections. Kemp et al. introduce spatiotemporal decomposition methods for weather-radar-based time series, thereby improving the detection of migratory dynamics by analyzing diel activity cycles and directional flow \cite{kemp2020spatiotemporal}. Such signal-processing techniques reduce false positives and improve the biological selectivity of radar-generated descriptors.
\subsubsection{Trajectory Reconstruction, Movement Modeling, and Forecasting}
Movement modeling beyond detection is crucial for predicting risk and grasping migration dynamics. Williams et al. proposed a method to predict bird movement using weather radar and environmental precursors, and showed that radar velocity fields can be incorporated into environmental prediction algorithms for hazard assessment \cite{williams2018prediction}. Specifically, their pipeline includes extraction of migration intensity, horizontal velocity vectors, altitude-stratified flow fields, and temporal smoothing filters—all directly relevant to bat migration risk prediction. Finally, Smith et al. expand these methods to bats in particular, demonstrating how enhanced radar processing, more streamlined clutter suppression, and multi-radar integration can be used to augment predictions of bat migratory behavior near wind-energy facilities \cite{smith2020batmigration}. Their work highlights species-specific radar signatures (e.g., emergence timing, reflectivity structure), multi-sensor fusion, and the utility of coupling radar outputs to ecological insights. This work defines a methodological framework for bat-specific radar monitoring that comprises weather-radar bioscatter extraction, micro-Doppler-based behavioral inference, classification of bats by polarimetric and Doppler features, and prediction of bat behavior using spatiotemporal radar models.

\subsection{Role of Radar in Bat Monitoring Systems}
Radar is an effective, reliable way to monitor bats and other flying animals, given the limited acoustic and visual tools available for this purpose. Because radar can detect movement not only in the dark but also under poor weather and over long distances, it fills significant gaps left by traditional sensing methods. Radar’s basic physics, which measures range, direction, and velocity from returned electromagnetic waves, makes it ideally suited for tracking fast-moving and hard-to-observe animals such as bats. Radar systems expanded from rudimentary pulse and Doppler radars during their formative years to high-frequency micro-Doppler units and nationwide weather-radar systems. These technologies enable researchers to record anything from individual bat flights at wind turbines to mass nightly migrations. The new wave of signal processing, clutter removal, and classification algorithms enables filtering the collected radar data, their clean interpretation, and their integration with other sensing devices—such as acoustic or visually recorded ones—which can generate highly valuable ecological information. It is important to note that radar is a powerful tool for bat monitoring because it can measure multiple dimensions of activity, performs well in challenging environments, and can be easily combined with modern analytical methodologies. While radar hardware, software, and analytics are well advanced, radar is becoming increasingly critical to the design of biological radar-based wildlife monitoring and conservation. 
\section*{Conclusion}
Fixed, unobtrusive sensing systems have played an important role in advancing global, long-term observation of bats and other volant wildlife in large-scale surveillance. Among the sensing modalities discussed—acoustic, camera-based computer vision, infrared, and radar—all have unique advantages, and their disadvantages are compensated for. Acoustic detection yields a great deal of behavioural and species-level information, but has limitations regarding environmental variances and call-specific bias. Cameras and computer vision offer visual validation and can facilitate scalable automated analysis, but illumination, occlusion, and triggering constraints greatly limit their performance. Infrared sensing helps us monitor the sky for extended periods with minimal interference, but has limitations in environmental sensitivity and in accurately resolving groups of rapidly moving or clustered individuals. Radar systems combine long-range weather-independent monitoring with an ability to discern flight trajectories, movement intensity, and even wingbeat features, placing radar as a tool that, even in aeroecology today, holds out a window of hope. Collectively, existing bat monitoring findings suggest that no single mode will suffice on its own, and the most comprehensive ecological understanding is achieved through multimodal integration. The use of acoustic, optical, infrared, and radar measurements may attenuate species-specific detection bias, improve classification precision, enhance cross-validation, and generate comprehensive overviews of bat behavior, population behavior, and environmental changes. Recent developments in edge computing, machine learning, multimodal data, and radar-based biological signal processing will help solidify this integration. In the long term, the development of these unobtrusive sensing systems will underpin more scalable, accurate, and ecologically sustainable approaches for monitoring of bat populations in a world of accelerating environmental and anthropogenic change.

\renewcommand{\bibfont}{\scriptsize}

\bibliographystyle{ACM-Reference-Format}
\bibliography{references_Erwei,references_maatla,references_isaac,references}
\end{document}